\documentclass[sigconf]{acmart}

\usepackage{booktabs}          
\usepackage{multirow}          
\usepackage{subcaption}
\usepackage{xspace} 

\AtBeginDocument{%
  \providecommand\BibTeX{{%
    \normalfont B\kern-0.5em{\scshape i\kern-0.25em b}\kern-0.8em\TeX}}}

\setcopyright{acmlicensed}
\setcopyright{cc}
\copyrightyear{2025}
\acmYear{2025}
\acmDOI{}

\acmConference[Conference acronym 'XX]{Make sure to enter the correct
  conference title from your rights confirmation email}{June 03--05,
  2018}{Woodstock, NY}
\acmISBN{978-1-4503-XXXX-X/18/06}

\graphicspath{{./images/}} 

\newcommand{\modelname}{OneTrans\xspace}
\newcommand{\Model}{\textsc{\modelname}\xspace}

\begin{document}

\title{{\modelname}: Unified Feature Interaction and Sequence Modeling with One Transformer in Industrial Recommender}


\author{Zhaoqi Zhang}
\authornote{These authors contributed equally.}              
\affiliation{%
  \institution{Nanyang Technological University\\ByteDance}
  \city{Singapore}
  \country{Singapore}
}
\email{zhaoqi.zhang@bytedance.com}

\author{Haolei Pei}
\authornotemark[1]                                          
\affiliation{%
  \institution{ByteDance}
  \city{Singapore}
  \country{Singapore}
}
\email{haolei.pei@bytedance.com}

\author{Jun Guo}
\authornotemark[1]                                          
\affiliation{%
  \institution{ByteDance}
  \city{Singapore}
  \country{Singapore}
}
\email{jun.guo@bytedance.com}

\author{Tianyu Wang}
\affiliation{%
  \institution{ByteDance}
  \city{Singapore}
  \country{Singapore}
}
\email{tianyu.wang01@bytedance.com}

\author{Yufei Feng}
\affiliation{%
  \institution{ByteDance}
  \city{Hangzhou}
  \country{China}
}
\email{fengyihui@bytedance.com}

\author{Hui Sun}
\affiliation{%
  \institution{ByteDance}
  \city{Hangzhou}
  \country{China}
}
\email{sunhui.sunh@bytedance.com}

\author{Shaowei Liu}
\authornote{Corresponding author.}                          
\affiliation{%
  \institution{ByteDance}
  \city{Singapore}
  \country{Singapore}
}
\email{liushaowei.nphard@bytedance.com}

\author{Aixin Sun}
\authornotemark[2]                                          
\affiliation{%
  \institution{Nanyang Technological University}
  \city{Singapore}
  \country{Singapore}
}
\email{axsun@ntu.edu.sg}

\renewcommand{\shortauthors}{Zhang et al.}

\begin{abstract}
In recommendation systems, scaling up feature-interaction modules (e.g., Wukong, RankMixer) or user-behavior sequence modules (e.g., LONGER) has achieved notable success. However, these efforts typically proceed on separate tracks, which not only hinders bidirectional information exchange but also prevents unified optimization and scaling. In this paper, we propose \textsc{\Model}, a unified Transformer backbone that simultaneously performs user-behavior sequence modeling and feature interaction. \textsc{\Model} employs a unified tokenizer to convert both sequential and non-sequential attributes into a single token sequence. The stacked \textsc{\Model} blocks share parameters across similar sequential tokens while assigning token-specific parameters to non-sequential tokens. Through causal attention and cross-request KV caching, \textsc{\Model} enables precomputation and caching of intermediate representations, significantly reducing computational costs during both training and inference. Experimental results on industrial-scale datasets demonstrate that \textsc{\Model} scales efficiently with increasing parameters, consistently outperforms strong baselines, and yields a 5.68\% lift in per-user GMV in online A/B tests.
\end{abstract}




\maketitle

\begingroup
\renewcommand\thefootnote{} 
\footnotetext{Accepted at The Web Conference 2026 (WWW 2026). Camera-ready version forthcoming.}
\addtocounter{footnote}{-1} 
\endgroup

\section{Introduction}
\label{sec:intro}

\begin{figure}
  \centering
  \includegraphics[width=1.0\linewidth]
  {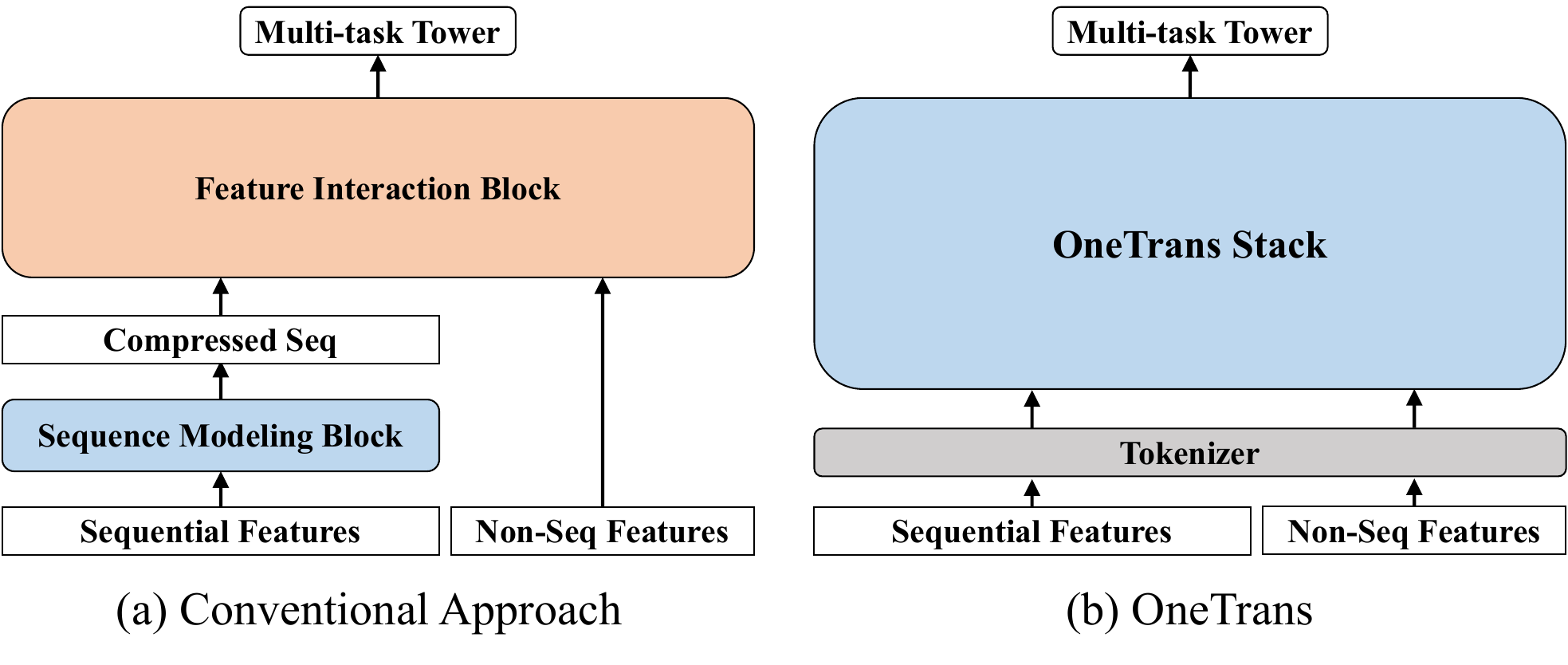} 
  \caption{Architectural comparison. \textbf{(a)} Conventional \emph{encode-then-interaction} pipeline encodes sequential features and merges non-sequential features before a post-hoc feature interaction block. \textbf{(b)} \textsc{\Model} performs joint modeling of both sequential and non-sequential features within a single \textsc{\Model} (Transformer-style)  stack.}
  \label{fig:arch-compare}
\end{figure}

Recommendation systems (RecSys) play a fundamental role in various information services, such as e-commerce~\cite{zhou2018deepnetworkclickthroughrate, feng2019deepsessionnetworkclickthrough}, streaming media~\cite{pancha2022pinnerformer, chang2023pepnet, xia2023transact} and social networks~\cite{zhang2024wukong, zhai2024actions}. 
Industrial RecSys generally adopt a cascaded ranking architecture~\cite{covington2016deep, liu2017cascade, qin2022rankflow}. 
First, a recall stage selects hundreds of candidates from billion-scale corpora~\cite{zhu2018learning, huang2024comprehensive}. Then, a ranking stage 
scores each candidate and returns the top-$k$ items~\cite{Wang_2021, gui2023hiformerheterogeneousfeatureinteractions, xia2023transact, zhang2024wukong, zhu2025rankmixerscalingrankingmodels}. 
Deep Learning Recommendation Models (DLRMs)~\cite{naumov2019deep} are widely adopted in the ranking stage of industrial recommenders.

We focus on the ranking stage in this paper, following the DLRM-style ranking paradigm.
For ranking, mainstream approaches iterate on two separate modules: 
(a) \emph{sequence modeling}, which encodes user multi-behavior sequences into candidate-aware representations using local attention or Transformer encoders~\cite{zhou2018deepnetworkclickthroughrate, kang2018selfattentivesequentialrecommendation, sun2019bert4recsequentialrecommendationbidirectional, chai2025longer}, 
and (b) \emph{feature interaction}, which learns high-order crosses among non-sequential features (e.g., user profile, item profile, and context) via factorization, explicit cross networks, or attention over feature groups~\cite{guo2018deepfmendtoendwide, Wang_2021, gui2023hiformerheterogeneousfeatureinteractions, zhu2025rankmixerscalingrankingmodels}. 
As shown in Fig.~\ref{fig:arch-compare}(a), these approaches typically encode user behaviors into a \emph{compressed} sequence representation, then concatenate it with non-sequential features and apply a feature-interaction module to learn higher-order interaction; we refer to this design as the \emph{encode-then-interaction} pipeline.

The success of large language models (LLMs) demonstrates that scaling model size (e.g., parameter size,  training data) yields predictable gains in performance~\cite{kaplan2020scaling}, inspiring similar investigations within RecSys~\cite{zhang2024wukong, chai2025longer, zhu2025rankmixerscalingrankingmodels}. 
For feature interaction, 
Wukong~\cite{zhang2024wukong} stacks Factorization Machine blocks with linear compression to capture high-order feature interactions and establishes scaling laws, while RankMixer~\cite{zhu2025rankmixerscalingrankingmodels} achieves favorable scaling through hardware-friendly token-mixing with token-specific feed-forward networks (FFNs).
For sequence modeling, LONGER\cite{chai2025longer} applies causal Transformers to long user histories and shows that scaling depth and width yields monotonic improvements. 
Although effective in practice, separating sequence modeling and feature interaction as independent modules introduces two major limitations. 
First, the encode-then-interaction pipeline restricts bidirectional information flow, limiting how static/context features shape sequence representations~\cite{zeng2024interformer}.
Second, module separation fragments execution and increases latency, whereas a single Transformer-style backbone can reuse LLM optimizations e.g., KV caching, memory-efficient attention, and mixed precision, for more effective scaling~\cite{gui2023hiformerheterogeneousfeatureinteractions}.
%

In this paper, we propose \textbf{\textsc{\Model}}, an innovative architectural paradigm with a unified Transformer backbone that jointly performs user-behavior sequence modeling and feature interaction. 
As shown in Fig.~\ref{fig:arch-compare}(b), \Model enables bidirectional information exchange within the unified backbone. It employs a unified tokenizer that converts both \emph{sequential} features (diverse behavior sequences) and \emph{non-sequential} features (static user/item and contextual features) into a single token sequence, which is then processed by a pyramid of stacked {\Model} blocks, a Transformer variant tailored for industrial RecSys. 
To accommodate the diverse token sources in RecSys, unlike the text-only tokens in LLMs, each \textsc{\Model} block adopts a \emph{mixed} parameterization similar to HiFormer~\cite{gui2023hiformerheterogeneousfeatureinteractions}. 
Specifically, all \emph{sequential} tokens (from sequential features) share a single set of Q/K/V and FFN weights, while each \emph{non-sequential} token (from non-sequential features) receives \emph{token-specific} parameters to preserve its distinct semantics.

Unlike conventional encode-then-interaction frameworks, \textsc{\Model} eliminates the architectural barrier between \emph{sequential} and \emph{non-sequential} features through a unified causal Transformer backbone. 
This formulation brings RecSys scaling in line with LLM practices: the \emph{entire} model can be scaled by adjusting backbone depth and width, while seamlessly inheriting mature LLM optimizations, such as FlashAttention~\cite{dao2022flashattention}, and mixed precision training~\cite{micikevicius2017mixed}.
Particularly, cross-candidate and cross-request KV caching~\cite{chai2025longer} reduces the time complexity from $O(C)$ to $O(1)$ for sessions with $C$ candidates, making large-scale {\Model} deployment feasible.

In summary, our main contributions are fourfold:
\textbf{(1) Unified framework.} We present \textsc{\Model}, a single Transformer backbone for ranking, equipped with a \emph{unified tokenizer} that encodes sequential and non-sequential features into one token sequence, and a \emph{unified Transformer block} that jointly performs sequence modeling and feature interaction.
\textbf{(2) Customization for recommenders.} To bridge the gap between LLMs and RecSys tasks, \textsc{\Model} introduces a \emph{mixed parameterization} that allocates token-specific parameters to diverse non-sequential tokens while sharing parameters for all sequential tokens.
\textbf{(3) Efficient training and serving.} We improve efficiency with a \emph{pyramid strategy} that progressively prunes sequential tokens and a \emph{cross-request KV Caching} that reuses user-side computations across candidates. In addition, we adopt LLM optimizations such as FlashAttention, mixed-precision training, and half-precision inference to further reduce memory and compute.
\textbf{(4) Scaling and deployment.} \textsc{\Model} demonstrates near log-linear performance gains with increased model size, providing evidence of a scaling law in real production data. When deployed online, it achieves statistically significant lifts on business KPIs while maintaining production-grade latency.

\section{Related Work}
\label{sec:related}

\begin{figure*}
\centering
\includegraphics[scale=0.8, clip]{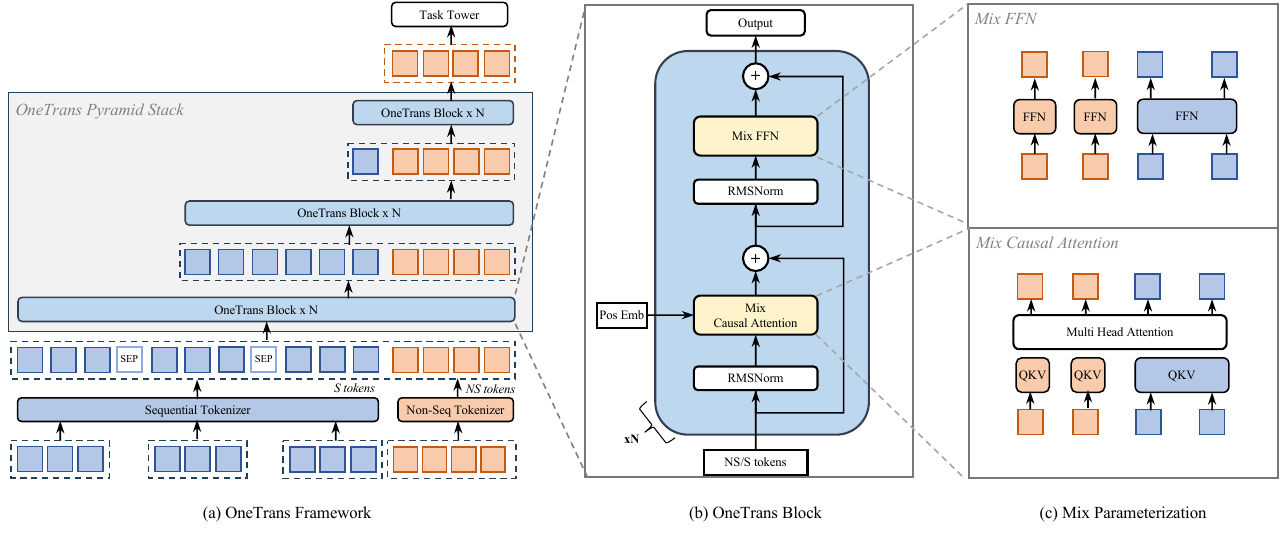}
\caption{\textbf{System Architecture.}
\textbf{(a)} \textsc{\Model} overview. Sequential (S, blue) and non-sequential (NS, orange) features are tokenized separately. After inserting \texttt{[SEP]} between user behavior sequences, the unified token sequence is fed into stacked \emph{\textsc{\Model} Pyramid Blocks} that progressively shrink the token length until it matches the number of NS tokens.
\textbf{(b)} \textsc{\Model} Block: a causal pre-norm Transformer Block with RMSNorm, \emph{Mixed Causal Attention} and \emph{Mixed FFN}.
\textbf{(c)} ``Mixed'' = mixed parameterization: S tokens share one set of QKV/FFN weights, while each NS token receives its own token-specific QKV/FFN.}
\label{fig:onetrans}
\end{figure*}

Early RecSys like DIN~\cite{zhou2018deepnetworkclickthroughrate} and its session-aware variants (DSIN)~\cite{feng2019deepsessionnetworkclickthrough} use local attention to learn candidate-conditioned summaries of user histories, but compress behaviors into fixed-length vectors per candidate, limiting long-range dependency modeling~\cite{zhou2018deepevolutionnetworkclickthrough}. 
Self-attentive methods like SASRec~\cite{kang2018selfattentivesequentialrecommendation}, BERT4Rec~\cite{sun2019bert4recsequentialrecommendationbidirectional}, and BST~\cite{chen2019behaviorsequencetransformerecommerce} eliminate this bottleneck by letting each position attend over the full history and improve sample efficiency with bidirectional masking. 
Recently, as scaling laws~\cite{kaplan2020scaling} in RecSys are increasingly explored, LONGER~\cite{chai2025longer} pushes sequence modeling toward industrial scales by targeting ultra-long behavioral histories with efficient attention and serving-friendly designs.
However, in mainstream pipelines these sequence encoders typically remain \emph{separate} from the feature-interaction stack, leading to late fusion rather than joint optimization with static contextual features~\cite{zeng2024interformer}.

On the feature-interaction side, early RecSys rely on manually engineered cross-features or automatic multiplicative interaction layers.
Classical models such as Wide{\allowbreak}\&{\allowbreak}Deep~\cite{cheng2016widedeeplearning}, FM/DeepFM~\cite{chang2010training,guo2018deepfmendtoendwide}, and DCN/DCNv2~\cite{wang2017deepcrossnetwork,Wang_2021} provide efficient low-order or bounded-degree interactions. However, as recent scaling studies observe~\cite{zhang2024wukong}, once the model stacks enough cross layers, adding more stops helping: model quality plateaus instead of continuing to improve.
To overcome the rigidity of preset cross forms, attention-based approaches automatically learn high-order interactions.
AutoInt~\cite{Song_2019} learns arbitrary-order relations, and HiFormer~\cite{gui2023hiformerheterogeneousfeatureinteractions} introduces group-specific projections to better capture heterogeneous, asymmetric interactions. 
With scaling up increasingly applied to \emph{feature-interaction} modules, large-scale systems such as Wukong~\cite{zhang2024wukong} demonstrate predictable gains by stacking FM-style interaction blocks with linear compression, 
while RankMixer~\cite{zhu2025rankmixerscalingrankingmodels} achieves favorable scaling via parallel token mixing and sparse MoE under strict latency budgets. 
However, these interaction modules typically adhere to the \emph{interaction} paradigm, which pushes interactions to a separate stage and blocks unified optimization with user sequence modeling~\cite{zeng2024interformer}.

To date, progress in RecSys has largely advanced along two independent tracks: sequence modeling and feature interaction. 
InterFormer~\cite{zeng2024interformer} attempts to bridge this gap through a summary-based bidirectional cross architecture that enables mutual signal exchange between the two components. However, it still maintains them as separate modules, and the cross architecture introduces both architectural complexity and fragmented execution. Without a unified backbone for joint modeling and optimization, scaling the system as an integrated whole remains challenging.
%

Recent work on Generative Recommenders (GRs) frames recommendation as sequential transduction and proposes efficient long-context backbones such as HSTU~\cite{zhai2024actions}. This line is complementary to DLRMs that rely on rich non-sequential (NS) features.

\section{Methodology}
\label{sec:method}

Before detailing our method, we briefly describe the task setting. 
In a cascaded industrial RecSys, each time the recall stage returns a candidate set (typically hundreds of candidate items) for a user $u$. The ranking model then predicts a score to each candidate item $i$:
\begin{equation}
\hat{y}_{u,i}
= f\left(i \,\big|\, \mathcal{NS},\mathcal{S};\Theta\right)
\label{eqn:task_def}
\end{equation}
where $\mathcal{NS}$ is a set of non-sequential features derived from the user, the candidate item, and the context; $\mathcal{S}$ is a set of historical behavior sequences from the user; and $\Theta$ are trainable parameters. Common task predictions include the click-through rate (CTR) and the post-click conversion rate (CVR).
\begin{equation}
\begin{aligned}
\text{CTR}_{u,i}&=P~\big(\text{click}=1\,\big|\,\mathcal{NS},\mathcal{S};\Theta\big),\\
\text{CVR}_{u,i}&=P~\big(\text{conv}=1\,\big|\,\text{click}=1,\mathcal{NS},\mathcal{S};\Theta\big).
\end{aligned}
\label{eqn:ctr_cvr_def}
\end{equation}

\subsection{\textsc{\Model} Framework Overview}
\label{sec:framework}
%
As illustrated in Fig.~\ref{fig:onetrans}(a), {\Model} employs a \emph{unified tokenizer} that maps sequential features $\mathcal{S}$ to S-tokens, and non-sequential features  $\mathcal{NS}$ to NS-tokens. A \emph{pyramid-stacked Transformer} then consumes the unified token sequence jointly within a single computation graph.
We denote the initial token sequence as
\begin{equation}
\mathbf{X}^{(0)} = \big[ \text{S-tokens} \,;\, \text{NS-tokens} \big] \in \mathbb{R}^{(L_{\text{S}} + L_{\text{NS}}) \times d}.
\end{equation}
This token sequence is constructed by concatenating $L_{\mathrm{S}}$ number of S-tokens
and $L_{\mathrm{NS}}$ number of NS-tokens, with all tokens having dimensionality $d$. Note that, the S-tokens contain learnable \texttt{[SEP]} tokens inserted to delimit boundaries between different kind of user-behavior sequences.
As shown in Fig.~\ref{fig:onetrans}(b), each {\Model} block progressively refines the token states through:
\begin{align}
\mathbf{Z}^{(n)} &= \mathrm{MixedMHA}\!\left(\mathrm{Norm}\big(\mathbf{X}^{(n-1)}\big)\right) + \mathbf{X}^{(n-1)}, \\
\mathbf{X}^{(n)} &= \mathrm{MixedFFN}\!\left(\mathrm{Norm}\big(\mathbf{Z}^{(n)}\big)\right) + \mathbf{Z}^{(n)}.
\end{align}
Here, $\mathrm{MixedMHA}$ (Mixed Multi-Head Attention) and $\mathrm{MixedFFN}$ (Mixed Feed-Forward Network) adopt a mixed parameterization strategy (see Fig.~\ref{fig:onetrans}(c)) sharing weights across sequential tokens, while assigning separate parameters to non-sequential tokens in both the attention and feed-forward layers. 

A unified causal mask enforces autoregressive constraints, restricting each position to attend only to preceding tokens. Specifically, NS-tokens are permitted to attend over the entire history of S-tokens, thereby enabling comprehensive cross-token interaction. By stacking such blocks with pyramid-style tail truncation applied to S-tokens, the model progressively distills compact high-order information into the NS-tokens. The final token states are then passed to task-specific heads for prediction.

By unifying non-sequential and sequential features into a unified token sequence and modeling them with a causal Transformer, \textsc{\Model} departs from the conventional \emph{encode-then-interaction} pipeline. This unified design naturally enables 
(i) \emph{intra-sequence} interactions within each behavior sequence, 
(ii) \emph{cross-sequence} interactions across multiple sequences, 
(iii) \emph{multi-source feature} interactions among item, user, and contextual features, and (iv) \emph{sequence-feature} interactions, \emph{all within a single Transformer stack}. 

The unified formulation enables us to seamlessly inherit mature LLM engineering optimizations, including KV caching and memory-efficient attention, thereby substantially reducing inference latency. We argue this unified formulation is well suited to tackling multi-sequence and cross-domain recommendation challenges in a single, and scalable architecture. Next, we detail the design.

\subsection{Features and Tokenization}
%
To construct the initial token sequence $\mathbf{X}^{(0)}$, {\Model} first applies a feature preprocessing pipeline that maps all raw feature inputs into embedding vectors. These embeddings are then partitioned into (i) a multi-behavior \emph{sequential} subset and (ii) a \emph{non-sequential} subset representing user, item, or context features. Separate tokenizers are applied to each subset.

\subsubsection{Non-Sequential Tokenization}
%
Non-sequential features $\mathcal{NS}$ include both numerical inputs (e.g., price, CTR) and categorical inputs (e.g., user ID, item category). All features are either bucketized or one-hot encoded and then embedded. Since industrial systems typically involve hundreds of features with varying importance, there are two options for controlling the number of non-sequential tokens, denoted by $L_{NS}$:

\medskip\noindent\textbf{Group-wise Tokenizer} \,(aligned with RankMixer~\cite{zhu2025rankmixerscalingrankingmodels}).\; 
Features are manually partitioned into semantic groups $\{ \mathbf{g}_1,\dots, \mathbf{g}_{L_{NS}} \}$. Each group is concatenated and passed through a group-specific MLP:
\begin{equation}
\text{NS-tokens} =
\big[\,\text{MLP}_1(\text{concat}(\mathbf{g}_1)),\dots,
\text{MLP}_{L_{NS}}(\text{concat}(\mathbf{g}_{L_{NS}}))\,\big].
\end{equation}

\medskip\noindent\textbf{Auto-Split Tokenizer}. 
Alternatively, all features are concatenated and projected once by a single MLP, then split:
\begin{equation}
\text{NS-tokens} =
\text{split}\Big(\text{MLP}(\text{concat}(\mathcal{NS})),\,L_{NS}\Big).
\end{equation}
Auto-Split Tokenizer reduces kernel launch overhead compared with Group-wise approach, by using a single dense projection. We will evaluate both choices through experiments. 

Ultimately, non-sequential tokenization yields $L_{NS}$ number of non-sequential tokens, each of dimensionality $d$. 

\subsubsection{Sequential Tokenization}

\Model accepts multi-behavior sequences as
\begin{equation}
\mathcal{S} = \{ \mathbf{S}_1, \dots, \mathbf{S}_n \}, \quad 
\mathbf{S}_i = \big[ \mathbf{e}_{i1}, \dots, \mathbf{e}_{iL_i} \big].
\end{equation}
Each sequence $\mathbf{S}_i$ consists of $L_i$ number of event embeddings $\mathbf{e}$, which is constructed by concatenating the item ID with its corresponding side information like  item category and price.

Multi-behavior sequences can vary in their raw dimensionality. Hence, for each sequence $\mathbf{S}_i$, we use one shared projection $\mathrm{MLP}_i$ to convert its all event $\mathbf{e}_{ij}$ as a common dimensionality $d$:
\begin{equation}
\tilde{\mathbf{S}}_i =
\big[\, \mathrm{MLP}_i(\mathbf{e}_{i1}), \dots, \mathrm{MLP}_i(\mathbf{e}_{iL_i}) \,\big]
\in \mathbb{R}^{L_i \times d}.
\end{equation}
Aligned sequences $\tilde{\mathbf{S}}_i$ are merged into a single token sequence by one of two rules: 1) \emph{Timestamp-aware}: interleave all events by time, with sequence-type indicators; 2) \emph{Timestamp-agnostic}: concatenate sequences by event impact, e.g.,  purchase $\rightarrow$ add-to-cart $\rightarrow$ click, inserting learnable \texttt{[SEP]} tokens between sequences. In the latter, behaviors with higher user intent are placed earlier in the sequence.
Ablation results indicate that, when timestamps are available, the timestamp-aware rule outperforms the impact-ordered alternative.
Formally, we have:
\begin{equation}
\text{S-Tokens} = \mathrm{Merge}\big(\tilde{\mathbf{S}}_1, \dots, \tilde{\mathbf{S}}_n\big) \in \mathbb{R}^{L_S \times d}, \quad
L_S = \sum_{i=1}^n L_i + L_\text{SEP}.
\end{equation}

\subsection{{\Model} Block}
\label{sec:one_block}
As shown in Fig.~\ref{fig:onetrans}(b), each {\Model} block is a pre-norm causal Transformer applied to a \emph{normalized} token sequence: $L_S$ \emph{sequential} S-tokens, followed by $L_{NS}$ \emph{non-sequential} NS-tokens. 
Inspired by the findings on heterogeneous feature groups~\cite{gui2023hiformerheterogeneousfeatureinteractions}, we make a lightweight modification to Transformer to allow a mixed parameter scheme, see Fig.~\ref{fig:onetrans}(c). Specifically, homogeneous S-tokens share one set of parameters. The NS-tokens, being heterogeneous across sources/semantics, receive token-specific parameters.

Unlike LLM inputs, the token sequence in RecSys combines sequential S-tokens with diverse NS-tokens whose value ranges and statistics differ substantially. Post-norm setups can cause attention collapse and training instability due to these discrepancies. To prevent this, we apply RMSNorm~\cite{zhang2019root} as pre-norm to \emph{all} tokens, aligning scales across token types and stabilizing optimization.

\subsubsection{Mixed (shared/token-specific) Causal Attention}
\Model adopts a standard multi-head attention (MHA) with a causal attention mask; the only change is how Q/K/V are parameterized.
Let \(\mathbf{x}_i\in\mathbb{R}^{d}\) be the \(i\)-th token.
To compute Q/K/V, we use a \emph{shared} projection for S-tokens (\(i\le L_S\)) and $L_{NS}$ \emph{token-specific} projections for NS-tokens (\(i > L_S\)):
\begin{equation}
\big(\mathbf{q}_i, \mathbf{k}_i, \mathbf{v}_i\big) \;=\; \big(\mathbf{W}^{Q}_i \mathbf{x}_i , \;\mathbf{W}^{K}_i \mathbf{x}_i , \; \mathbf{W}^{V}_i \mathbf{x}_i \big),
\label{eqn:per_token_qkv}
\end{equation}
where $\mathbf{W}^{\Psi}_i$ ($\Psi \in \{Q,K,V\}$) follows a mixed parameterization scheme:
\begin{equation}
\mathbf{W}^{\Psi}_i \;=\;
\begin{cases}
\mathbf{W}^{\Psi}_{\mathrm{S}}, & i \le L_S \quad (\text{shared for S-tokens}),\\[1.5mm]
\mathbf{W}^{\Psi}_{\mathrm{NS},i}, & i > L_S \quad (\text{token-specific for NS-tokens}).
\end{cases}
\label{eqn:per_token}
\end{equation}

Attention uses a standard \emph{causal} mask, with NS-tokens placed \emph{after} S-tokens. This induces:
(1) \textbf{S-side.} Each S-token attends only to earlier $S$ positions. For \emph{timestamp-aware} sequences, every event conditions on its history; for \emph{timestamp-agnostic} sequences (ordered by intent, e.g., purchase $\rightarrow$ add-to-cart $\rightarrow$ click/impression), causal masking lets high-intent signals inform and filter later low-intent behaviors.
(2) \textbf{NS-side.} Every NS-token attends to the \emph{entire} $S$ history, effectively a target-attention aggregation of sequence evidence, and to \emph{preceding} NS-tokens, increasing token-level interaction diversity.
(3) \textbf{Pyramid support.} On both S and NS sides, causal masking progressively concentrates information toward later positions, naturally supporting the pyramid schedule that prunes tokens layer by layer, to be detailed shortly.

\subsubsection{Mixed (shared/token-specific) FFN}
Similarly, the feed-forward network follows the same parameterization strategy: token-specific FFNs for NS-tokens, and a shared FFN for S-tokens,
\begin{equation}
\mathrm{MixedFFN}(\mathbf{x}_i) = \mathbf{W}^{2}_i \, \phi(\mathbf{W}^{1}_i \mathbf{x}_i).
\end{equation}
Here \(\mathbf{W}^{1}_i\) and \(\mathbf{W}^{2}_i\) follow the mixed parameterization of Eqn.~(\ref{eqn:per_token}), i.e., shared for \(i\le L_S\) and token-specific for \(i>L_S\).

In summary, relative to a standard causal Transformer, {\Model} changes only the \emph{parameterization}: NS-tokens use \emph{token-specific} QKV and FFN; S-tokens \emph{share} a single set of parameters. A single causal mask ties the sequence together, allowing NS-tokens to aggregate the entire behavior history while preserving efficient, Transformer-style computation.

\subsection{Pyramid Stack}
\label{sec:pyramid_stack}

As noted in Section~\ref{sec:one_block}, causal masking concentrates information toward later positions.
Exploiting this recency structure, we adopt a \emph{pyramid} schedule: at each {\Model} block layer, only a subset of the most recent S-tokens issue queries, while keys/values are still computed over the full sequence; the query set shrinks with depth.

Let $\mathbf{X}=\{\mathbf{x}_i\}_{i=1}^{L}$ be the input token list and $\mathcal{Q}=\{L{-}L'{+}1,\dots,L\}$ denote a tail index set  with $L'\le L$.
Following Eqn.~\ref{eqn:per_token}, we modify \emph{queries} as $i\in\mathcal{Q}$:
\begin{align}
\mathbf{q}_i \;=\; \mathbf{W}^{Q}_i\,\mathbf{x}_i, \qquad i\in\mathcal{Q},
\end{align}
while keys and values are computed as usual over the full sequence $\{1,\dots,L\}$.
After attention, only outputs for $i\in\mathcal{Q}$ are retained, reducing the token length to $L'$ and forming a pyramidal hierarchy across layers.

This design yields two benefits:
(i) \emph{Progressive distillation}: long behavioral histories are funneled into a small tail of queries, focusing capacity on the most informative events and consolidating information into the NS-tokens; and
(ii) \emph{Compute efficiency}: attention cost becomes $O\big(LL'd\big)$ and FFN scales linearly with $L'$. Shrinking the query set directly reduces FLOPs and activation memory.

\subsection{Training and Deployment Optimization}
\label{subsec:serving-opt}

\subsubsection{Cross Request KV Caching}

In industrial RecSys, samples from the same request are processed contiguously both during training and serving: their S-tokens remain identical across candidates, while NS-tokens vary per candidate item. Leveraging this structure, we integrate the widely adopted KV Caching~\cite{chai2025longer} into {\Model}, yielding a unified two-stage paradigm.

\medskip\noindent\textbf{Stage I (S-side, once per request).} Process all S-tokens with causal masking and cache their key/value pairs and attention outputs. This stage executes \emph{once} per request.

\medskip\noindent\textbf{Stage II (NS-side, per candidate).} For each candidate, compute its NS-tokens and perform cross-attention against the cached S-side keys/values, followed by token-specific FFN layers. Specially, candidate-specific sequences (e.g., SIM~\cite{pi2020search}) are pre-aggregated into NS-tokens via pooling, as they cannot reuse the shared S-side cache.

The KV Caching amortizes S-side computation across candidates, keeping per-candidate work lightweight and eliminating redundant computations for substantial throughput gains.

Since user behavioral sequences are append-only, we extend KV Caching \emph{across requests}: each new request reuses the previous cache and computes only the incremental keys/values for newly added behaviors. 
This reduces per-request sequence computation from $O(L)$ to $O(\Delta L)$, where $\Delta L$ is the number of new behaviors since the last request.

\subsubsection{Unified LLM Optimizations.}
\label{subsec:FlashAttention.}

We employ FlashAttention-2~\cite{dao2023flashattention2} to reduce attention I/O and the quadratic activation footprint of vanilla attention via tiling and kernel fusion, yielding lower memory usage and higher throughput in both training and inference. 
To further ease memory pressure, we use mixed-precision training (BF16/FP16)~\cite{micikevicius2018mixed} together with activation recomputation~\cite{gruslys2016memory}, which discards selected forward activations and recomputes them during backpropagation. 
This combination trades modest extra compute for substantial memory savings, enabling larger batches and deeper models without architectural changes.

\section{Experiments}
\label{sec:exp}

Through both offline evaluations and online tests, we aim to answer the following Research Questions (RQs):
\textbf{RQ1: Unified stack vs.\ encode--then--interaction.} 
Does the \emph{single Transformer stack} yield consistent performance gains under the comparable compute?
\textbf{RQ2: Which design choices matter?} 
We conduct ablations on the \emph{input layer} (e.g., tokenizer, sequence fusion) and the \emph{\Model block} (e.g., parameter sharing, attention type, pyramid stacking) to evaluate the importance of different design choices for performance and efficiency.
\textbf{RQ3: Systems efficiency.} Do pyramid stacking, cross-request KV Caching, FlashAttention-2, and mixed precision with recomputation reduce FLOPs/memory and latency under the same {\Model} graph?
\textbf{RQ4: Scaling law.} 
As we scale length (token sequence length), width ($d_{\text{model}}$), depth (number of layers), do loss/performance exhibit the expected \emph{log-linear} trend?
\textbf{RQ5: Online A/B Tests.} Does deploying {\Model} online yield statistically significant lifts in key business metrics (e.g., order/u, GMV/u) under production latency constraints?

\subsection{Experimental Setup}
\label{ssec:expSetup}

\subsubsection{Dataset}

\begin{table}[t]
\centering
\caption{Dataset overview for {\Model} experiments.}
\label{tab:dataset_overview}
\begin{tabular}{lr}
\toprule
Metric & Value \\
\midrule
\# Impressions (samples) & 29.1B \\
\# Users (unique) & 27.9M \\
\# Items (unique) & 10.2M \\
Daily impressions (mean $\pm$ std) & 118.2M $\pm$ 14.3M \\
Daily active users (mean $\pm$ std) & 2.3M $\pm$ 0.3M \\
\bottomrule
\end{tabular}
\end{table}

For offline evaluation, we evaluate {\Model} in a large-scale industrial ranking scenario using production logs under strict privacy compliance (all personally identifiable information is anonymized and hashed). Data are split chronologically, with all features snapshotted at impression time to prevent temporal leakage and ensure online-offline consistency. Labels (e.g., clicks and orders) are aggregated within fixed windows aligned with production settings. Table~\ref{tab:dataset_overview} summarizes the dataset statistics.

\begin{table*}[t]
\centering
\caption{Offline effectiveness (CTR/CVR) and efficiency; higher AUC/UAUC is better. 
* indicates models deployed in our production in chronological order: 
\emph{DCNv2+DIN} $\rightarrow$ \emph{RankMixer+DIN}  $\rightarrow$ \emph{RankMixer+Transformer} 
$\rightarrow$ \emph{\textsc{\Model}\textsubscript{S}} $\rightarrow$ 
\emph{\textsc{\Model}\textsubscript{L}}}
\label{tab:rq1_all}
\begin{tabular}{ll|cc|cc|rr}
\toprule
\multirow{2}{*}{\textbf{Type}} &\multirow{2}{*}{Model} & \multicolumn{2}{c|}{CTR} & \multicolumn{2}{c|}{CVR (order)} & \multicolumn{2}{c}{Efficiency} \\
\cmidrule(lr){3-4}\cmidrule(lr){5-6}\cmidrule(lr){7-8}
& & AUC\,$\uparrow$ & UAUC\,$\uparrow$ & AUC\,$\uparrow$ & UAUC\,$\uparrow$ & Params (M) & TFLOPs \\
\midrule
\textbf{(1) Base model} &DCNv2 + DIN (base)*      & 0.79623 & 0.71927 & 0.90361 & 0.71955 & 10  & 0.06 \\
\midrule
\multirow{3}{*}{\textbf{(2) Feature-interaction}}& Wukong + DIN                      & +0.08\% & +0.11\% & +0.14\% & +0.11\% & 28  & 0.54 \\
&HiFormer + DIN                    & +0.11\% & +0.18\% & +0.23\% & -0.20\% & 108 & 1.35 \\
&RankMixer + DIN*                  & +0.27\% & +0.36\% & +0.43\% & +0.19\% & 107 & 1.31 \\
\midrule
\multirow{3}{*}{\textbf{(3) Sequence-modeling}}&RankMixer + StackDIN              & +0.40\% & +0.37\% & +0.63\% & -1.28\% & 108 & 1.43 \\
&RankMixer + LONGER                & +0.49\% & +0.59\% & +0.47\% & +0.44\% & 109 & 1.87 \\
&RankMixer + Transformer*          & +0.57\% & +0.90\% & +0.52\% & +0.75\% & 109 & 2.51 \\
\midrule
\multirow{2}{*}{\textbf{(4) Unified framework}}&\textsc{\Model}\textsubscript{S}*           & +1.13\% & +1.77\% & +0.90\% & +1.66\% & 91  & 2.64 \\
&\textsc{\Model}\textsubscript{L} (default)* & +1.53\% & +2.79\% & +1.14\% & +3.23\% & 330 & 8.62 \\
\bottomrule
\end{tabular}
\end{table*}

\subsubsection{Tasks and Metrics}
We evaluate two binary ranking tasks as defined in Eqn.~(\ref{eqn:ctr_cvr_def}): CTR and CVR. Performance is measured by AUC and UAUC (impression-weighted user-level AUC).

\textbf{Next-batch evaluation.}
Data are processed chronologically. For each mini-batch, we (i) log predictions in eval mode, then (ii) train on the same batch. AUC and UAUC are computed daily from each day's predictions and finally macro-averaged across days.

\textbf{Efficiency metrics.}
We report \emph{Params} (model parameters excluding sparse embeddings) and \emph{TFLOPs} (training compute in TFLOPs at batch size 2048).
\subsubsection{Baselines}
We construct industry-standard model combinations as baselines using the same features and matched compute budgets. Under the \emph{encode-then-interaction} paradigm, we start from the widely-used production baseline \textbf{DCNv2+DIN}~\cite{Wang_2021,zhou2018deepnetworkclickthroughrate} and progressively strengthen the \textbf{feature-interaction} module: DCNv2  $\rightarrow$ Wukong~\cite{zhang2024wukong} $\rightarrow$ HiFormer~\cite{gui2023hiformerheterogeneousfeatureinteractions} $\rightarrow$ RankMixer~\cite{zhu2025rankmixerscalingrankingmodels}. 
With RankMixer fixed, we then vary the \textbf{sequence-modeling} module: StackDIN $\rightarrow$ Transformer~\cite{chen2019behaviorsequencetransformerecommerce} $\rightarrow$ LONGER~\cite{chai2025longer}.

\subsubsection{Hyperparameter Settings.}
\label{sec:hparam}
We report two settings: 
\textbf{\textsc{\Model}\textsubscript{S}} uses \emph{6} stacked {\Model} blocks width \(d{=}256\), and \(H{=}4\) heads, targeting \(\approx\!100\)M parameters.
\textbf{\textsc{\Model}\textsubscript{L}} scales to \emph{8} layers with width \(d{=}384\).
%
%
%


Inputs are processed through a unified tokenizer (timestamp-aware fusion for multi-behavior sequences; \emph{Auto-Split} for non-sequential features) and a heuristic pyramid schedule that, at each layer, linearly shrinks the number of sequential \emph{query} tokens from 1190 to 12 (\textsc{\Model}\textsubscript{S}) / from 1500 to 16 (\textsc{\Model}\textsubscript{L}). Concretely, we linearly reduce the number of sequential \emph{query} tokens across layers, rounding the token count at each layer to the nearest multiple of 32, and set the top layer to match the number of non-sequential tokens.

\textbf{Optimization and infrastructure.}
We use a dual-optimizer strategy without weight decay: sparse embeddings are optimized with Adagrad ($\beta_1{=}0.1$, $\beta_2{=}1.0$), and dense parameters with RMSProp (lr${=}0.005$, alpha${=}0.99999$,momentum${=}0$).
We apply stabilization techniques commonly used in large-scale Transformer training, including Pre-Norm~\cite{xiong2020layer}, and global grad-norm clipping~\cite{shoeybi2019megatron}.
The per-GPU batch size is set to 2048 during training, with gradient clipping thresholds of 90 for dense layers and 120 for sparse layers. For online inference, we adopt a smaller batch size of 100 per GPU to balance throughput and latency.
Training uses data-parallel all-reduce on 16 H100 GPUs.

\subsection{RQ1: Performance Evaluation}
\label{sec:perf_rq1}

We anchor our comparison on DCNv2+DIN, the pre-scaling production baseline in our scenario (Table~\ref{tab:rq1_all}). Under the \emph{encode-then-interaction} paradigm, scaling either component independently is beneficial: upgrading the \emph{feature interaction} module (DCNv2 $\rightarrow$ Wukong $\rightarrow$ HiFormer $\rightarrow$ RankMixer) or the \emph{sequence modeling} module (StackDIN $\rightarrow$ Transformer $\rightarrow$ LONGER) yields consistent gains in CTR AUC/UAUC and CVR AUC. In our system, improvements above $+0.1\%$ in these metrics are considered meaningful, while gains above $+0.3\%$ typically correspond to statistically significant effects in online A/B tests. However, CVR UAUC is treated cautiously due to smaller per-user sample sizes and higher volatility.

Moving to a unified design, \textsc{\Model}\textsubscript{S} surpasses the baseline by \(+1.13\%/+1.77\%\) (CTR AUC/UAUC) and \(+0.90\%/+1.66\%\) (CVR AUC/UAUC). 
At a comparable parameter scale, it also outperforms \text{RankMixer+Transformer} with similar training FLOPs (2.64T vs.\ 2.51T), demonstrating the benefits of unified modeling.
Scaling further, \textsc{\Model}\textsubscript{L} delivers the best overall improvement of \( +1.53\%\allowbreak /+2.79\% \) (CTR AUC/UAUC) and \( +1.14\%/+3.23\% \) (CVR AUC/UAUC), showing a predictable quality performance as model capacity grows.

In summary, unifying sequence modeling and feature interaction in a single Transformer yields more reliable and compute-efficient improvements than scaling either component independently.

\subsection{RQ2: Design Choices via Ablation Study}
\label{sec:ablation_rq2}

\begin{table*}
\centering
\caption{Impact of the choices of input design and \Model block design, using the \textbf{OneTrans\textsubscript{S}} model as the reference.}
\label{tab:rq2_unified}
\begin{tabular}{ll|cc|cc|rr}
\toprule
\multirow{2}{*}{\textbf{Type}} & \multirow{2}{*}{Variant} & \multicolumn{2}{c|}{CTR} & \multicolumn{2}{c|}{CVR (order)} & \multicolumn{2}{c}{Efficiency} \\
\cmidrule(lr){3-4}\cmidrule(lr){5-6}\cmidrule(lr){7-8}
& & AUC\,$\uparrow$ & UAUC\,$\uparrow$ & AUC\,$\uparrow$ & UAUC\,$\uparrow$ & Params (M) & TFLOPs \\
\midrule
\multirow{3}{*}{\textbf{Input}}
& Group-wise Tokenzier & -0.10\% & -0.30\% & -0.12\% & -0.10\% & 78 & 2.35 \\
& Timestamp-agnostic Fusion & -0.09\% & -0.22\% & -0.20\% & -0.21\% & 91 & 2.64 \\
& Timestamp-agnostic Fusion w/o Sep Tokens & -0.13\% & -0.32\% & -0.29\% & -0.33\% & 91 & 2.62 \\
\midrule
\multirow{3}{*}{\textbf{OneTrans Block}}
& Shared parameters & -0.15\% & -0.29\% & -0.14\% & -0.29\% & 24 & 2.64 \\
& Full attention & +0.00\% & +0.01\% & -0.03\% & +0.06\% & 91 & 2.64 \\
& w/o pyramid stack & -0.05\% & +0.06\% & -0.04\% & -0.42\% & 92 & 8.08 \\
\bottomrule
\end{tabular}
\end{table*}


\begin{figure}[t]
  \centering
    \begin{subfigure}[b]{0.35\textwidth}
        \includegraphics[trim={0.3cm 0.3cm 0.2cm 0.2cm},clip,width=\linewidth]{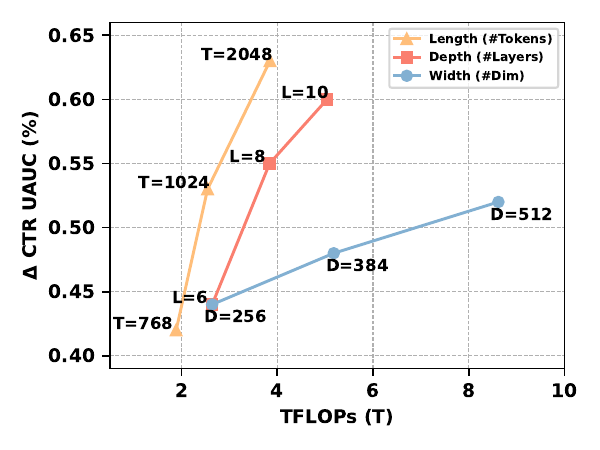}
        \caption{Trade-off: FLOPs vs.\ $\Delta$UAUC}
        \label{fig:scaling:a}
    \end{subfigure}
    \begin{subfigure}[b]{0.35\textwidth}
\includegraphics[trim={0.3cm 0.3cm 0.2cm 0.2cm},clip, width=\linewidth]{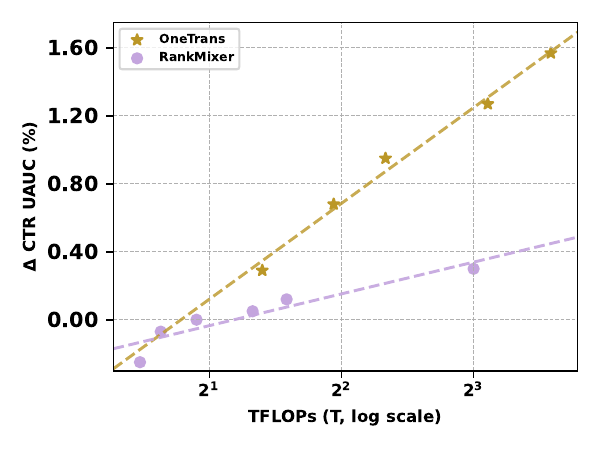}
        \caption{Scaling law: $\Delta$UAUC vs.\ FLOPs (log)}
        \label{fig:scaling:b}
    \end{subfigure}
   \caption{Comparison of trade-off and scaling law.}
  \label{fig:scaling}
\end{figure}

We perform an ablation study of the proposed \textsc{\Model} model to quantify the contribution of key design choices. The complete results are summarized in Table~\ref{tab:rq2_unified}. We evaluate the following variants: 
\textbf{Input variants:}
i) Replacing the \textit{Auto-Split Tokenizer} with a \textit{Group-wise Tokenizer} (Row~1);
ii) Using a timestamp-\textit{agnostic} fusion strategy instead of the timestamp-\textit{aware} sequence fusion (Row~2);
iii) Removing \texttt{[SEP]} tokens in the timestamp-\textit{aware} sequence fusion (Row~3);
\textbf{\Model block variants:}
i) Sharing a single set of Q/K/V and FFN parameters across \emph{all} tokens, instead of assigning separate parameters to NS-tokens (Row~4);
ii) Replacing causal attention with full attention (Row~5);
iii) Disabling the pyramid stack by keeping the full token sequence at \emph{all} layers (Row~6).

In summary, the ablations show that 
1) \textbf{Auto-Split Tokenizer} provides a clear advantage over manually grouping non-sequential features into tokens, indicating that allowing the model to automatically build non-sequential tokens is more effective than relying on human-defined feature grouping; 
2) \textbf{Timestamp-aware fusion} beats intent-based ordering when timestamps exist, suggesting that temporal ordering should be prioritized over event impact; 
3) Under timestamp-\emph{agnostic} fusion, learnable \textbf{\texttt{[SEP]} tokens} help the model separate sequences;
4) \textbf{Token-specific parameters} for NS-tokens outperform a shared projection, enabling better feature discrimination;
5) \textbf{Causal and full attention} perform similarly, but full attention disables standard optimizations such as KV caching;
6) Keeping full-length tokens at all layers provides no benefit: {\Model} effectively summarizes information into a small tail, so the \textbf{pyramid design} can safely prune queries to save computation. Moreover, under a fixed TFLOPs budget, the pyramid design supports \emph{close to} $1.75\times$ longer sequences than a full-length design, better exploiting gains from length extension.

\subsection{RQ3: Systems Efficiency}
\label{sec:rq4_efficiency}

\begin{table*}[t]
\centering
\caption{Impact of variants against the unoptimized \textsc{\Model}\textsubscript{S}. 
Memory is peak GPU usage.}
\label{tab:rq4_efficiency}
\begin{tabular}{l|rr|rr}
\toprule
\multirow{2}{*}{Variant} 
& \multicolumn{2}{c|}{\textbf{Training}} 
& \multicolumn{2}{c}{\textbf{Inference}} \\
\cmidrule(lr){2-3}\cmidrule(lr){4-5}
& Runtime (ms) & Memory (GB)
& Latency (p99; ms) & Memory (GB) \\
\midrule
Unoptimized \textsc{\Model}\textsubscript{S}             & 407    & 53.13  & 54.00  & 1.70 \\
+ Pyramid stack&                        \(-28.7\%\) & \(-42.6\%\) & \(-8.4\%\) & \(-6.9\%\) \\
+ Cross-Request KV Caching               & \(-30.2\%\) & \(-58.4\%\) & \(-29.6\%\) & \(-52.9\%\) \\
+ FlashAttention                      & \(-50.1\%\) & \(-58.9\%\) & \(-12.3\%\) & \(-11.6\%\) \\
+ Mixed Precision with Recomputation                    & \(-32.9\%\) & \(-49.0\%\) & \(-69.1\%\) & \(-30.0\%\) \\
\bottomrule
\end{tabular}%
\end{table*}

To quantify the optimizations in Section~\ref{subsec:serving-opt}, we ablate them on an unoptimized \textbf{\textsc{\Model}\textsubscript{S}} baseline and report training/inference metrics in Table~\ref{tab:rq4_efficiency}.

As shown in the table, (i) \textbf{Pyramid stack} reduces both training cost (runtime/memory) and serving overhead (p99 latency/memory) by pruning sequential \emph{query} tokens; (ii) \textbf{cross-request KV caching} removes redundant sequence computation, consistently improving runtime/latency and memory in both training and serving; (iii) \textbf{FlashAttention} yields substantial training gains with modest serving improvements; and (iv) \textbf{mixed precision with recomputation} provides the largest serving gains (p99 latency and inference memory), while also improving training efficiency.

These results demonstrate the effectiveness of LLM optimizations for large-scale recommendation. 
Building on these results, we scale to \textbf{\textsc{\Model}\textsubscript{L}} and show it maintains online efficiency comparable to the much smaller DCNv2+DIN baseline (Table~\ref{tab:baseline_half}), highlighting that a \emph{unified} Transformer backbone enables direct adoption of LLM optimizations.

\subsection{RQ4: Scaling-Law Validation} 
\label{sec:scaling_law}

\begin{table}[t]
\centering
\caption{Key efficiency comparison between \textsc{\Model}\textsubscript{L} and the DCNv2+DIN baseline.}
\label{tab:baseline_half}
\begin{tabular}{lcc}
\toprule
\textbf{Metric} & \textbf{DCNv2+DIN} & \textbf{\textsc{\Model}\textsubscript{L}} \\
\midrule
TFLOPs                        & 0.06  & 8.62 \\
Params (M)                          & 10    & 330 \\
MFU                             & 13.4  & 30.8 \\
Inference Latency (p99, ms)          & 13.6   & 13.2 \\
Training Memory (GB)            & 20     & 32  \\
Inference Memory (GB)           & 1.8    & 0.8 \\
\bottomrule
\end{tabular}
\end{table}

We probe \emph{scaling laws} for \textsc{\modelname} along three axes:
(1) \textbf{length} - input token sequence length,
(2) \textbf{depth} - number of stacked blocks, and
(3) \textbf{width} - hidden-state dimensionality.

As shown in Fig.~\ref{fig:scaling}(a), increasing \emph{length} yields the largest gains by introducing more behavioral evidence.
Between \emph{depth} and \emph{width}, we observe a clear trade-off: increasing \emph{depth} generally delivers larger performance improvements than simply widening \emph{width}, as deeper stacks extract higher-order interactions and richer abstractions.
However, deeper models also increase serial computation, whereas widening is more amenable to parallelism. Thus, choosing between \emph{depth} and \emph{width} should balance performance benefits against system efficiency under the target hardware budget.

We further analyze scaling-law behavior by jointly widening and deepening \textsc{\Model}, and --- for comparison --- by scaling the \textbf{RankMixer+Transformer} baseline on the RankMixer side till 1B; we then plot $\Delta$UAUC versus training FLOPs on a log scale. 
As shown in Fig.~\ref{fig:scaling}(b), \textsc{\Model} and RankMixer both exhibit clear log-linear trends, but \textsc{\Model} shows a \emph{steeper} slope, likely because RankMixer-centric scaling lacks a unified backbone and its MoE-based expansion predominantly widens the FFN hidden dimension.
Together, these results suggest that \textsc{\Model} is more \emph{parameter- and compute-efficient}, offering favorable performance--compute trade-offs for industrial deployment.

While we can deploy \textsc{\Model}\textsubscript{L} under strict online p99 latency constraints, scaling substantially beyond this regime remains constrained by online efficiency, and we leave further system–model co-optimizations to future work.

\subsection{RQ5: Online A/B Tests}
\label{sec:online_ab}

We assess the business impact of \textsc{\Model} in two large-scale industrial scenarios: (i) \emph{Feeds} (home feeds), and (ii) \emph{Mall} (the overall setting that includes Feeds and other sub-scenarios). 
Traffic is split at the user/account level with hashing and user-level randomization. Both the \emph{control} and \emph{treatment} models are trained and deployed with the past 1.5 years of production data to ensure a fair comparison. 

Our prior production baseline, \textbf{RankMixer+Transformer}, serves as the \emph{control} (\(\approx 100\)M neural-network parameters) and does not use sequence KV caching. The \emph{treatment} deploys {\Model}\textsubscript{L} with the serving optimizations described in Section~\ref{subsec:serving-opt}.

We report user-level \textbf{click/u}, \textbf{order/u}, and \textbf{gmv/u} as relative deltas (\(\Delta\%\)) versus the \emph{RankMixer+Transformer} control with two-sided 95\% CIs (user-level stratified bootstrap), and \textbf{end-to-end latency}, measured as the relative change in p99 per-impression time from request arrival to response emission (\(\Delta\%\); lower is better).
As shown in Table~\ref{tab:ab_results}, \textsc{\Model}\textsubscript{L} delivers consistent gains. In \emph{Feeds}, it achieves \(+7.737\%\) \texttt{click/u}, \(+4.3510\%\) \texttt{order/u}, \(+5.6848\%\) \texttt{gmv/u}, and \(-3.91\%\) latency. In \emph{Mall}, it achieves \(+5.143\%\) \texttt{click/u}, \(+2.5772\%\) \texttt{order/u}, \(+3.6696\%\) \texttt{gmv/u}, and \(-3.26\%\) latency. These results indicate that the unified modeling framework improves business metrics while reducing serving time relative to a strong non-unified baseline.

We further observe a \(+0.7478\%\) increase in user \emph{Active Days} and a significant improvement of \(+13.59\%\) in \emph{cold-start product order/u}, highlighting the strong generalization capability of the proposed model.


\begin{table}[t]
\centering
\caption{Online A/B results: \textsc{\Model}\textsubscript{L} (treatment) vs.\ RankMixer+Transformer (control).
Click/u, Order/u, GMV/u, are relative deltas (\%). Latency is the \emph{relative} end-to-end per-impression change \(\Delta\%\) (lower is better). * denotes $p<0.05$, and ** for $p<0.01$}
\label{tab:ab_results}
\begin{tabular}{l|ccccc}
\toprule
Scenario & click/u & order/u & gmv/u & Latency (\,\(p99\)\,)\,$\downarrow$ \\
\midrule
Feeds
& \(+7.737\%\)** 
& \(+4.351\%\)*
& \(+5.685\%\)*
& \(-3.91\%\) \\
Mall
& \(+5.143\%\)** 
& \(+2.577\%\)**
& \(+3.670\%\)*
& \(-3.26\%\) \\
\bottomrule
\end{tabular}
\end{table}

\section{Conclusion}

We present \textsc{\Model}, a unified Transformer backbone for personalized ranking to replace the conventional \emph{encode\textendash\allowbreak then\textendash\allowbreak interaction}. 
A unified tokenizer converts both sequential and non-sequential attributes into one token sequence, and a unified Transformer block jointly performs sequence modeling and feature interaction via shared parameters for homogeneous (sequential) tokens and token-specific parameters for heterogeneous (non-sequential) tokens. 
To make the unified stack efficient at scale, we adopt a pyramid schedule that progressively prunes sequential tokens and a cross-request KV Caching that reuses user-side computation; the design further benefits from LLM-style systems optimizations (e.g., FlashAttention, mixed precision). 
Across large-scale evaluations, \textsc{\Model} exhibits near log-linear performance gains as width/depth increase, and delivers statistically significant business lifts while maintaining production-grade latency. 
We believe this unified design offers a practical way to scale recommender systems while reusing the system optimizations that have powered recent LLM advances.

\bibliographystyle{ACM-Reference-Format}
\balance
\bibliography{references.bib}

\end{document}